\documentclass[11pt]{article}
\usepackage{amssymb}
\usepackage{amsmath}
\usepackage{amsfonts}
\usepackage{geometry}
\usepackage[onehalfspacing]{setspace}

\newtheorem{theorem}{Theorem}[section]

\newtheorem{assumption}{Assumption}[section]

\input{tcilatex}
\begin{document}

\title{\textbf{Analysis of Interactive Fixed Effects Dynamic\ Linear Panel
Regression with Measurement Error\thanks{This is the last working paper version of the paper published in \textit{Economics Letters} \textbf{117}(1), 239--242, 2012; doi:10.1016/j.econlet.2012.04.109. %
We thank Badi Baltagi and an anonymous referee for helpful comments and
suggestions. Moon thanks the NSF for financial support.}}}
\author{Nayoung Lee\thanks{%
Lee (Corresponding Author): Department of Economics, Chinese University of
Hong Kong; Shatin, New Territories, Hong Kong; Email:
nayoung.lee@cuhk.edu.hk; Telephone: (852) 3943-8004; Fax: (852) 2603-5805.
Moon: Department of Economics, University of Maryland, Tydings Hall, Room
3105, College Park, MD 20742; E-mail: moon@econ.umd.edu. Weidner: Department
of Economics, University College London, Gower St., London, WC1E 6BT, U.K.;
E-mail: m.weidner@ucl.ac.uk. } \\
CUHK \and Hyungsik Roger Moon \\
USC \& U of Maryland \and Martin Weidner \\
UCL and CeMMAP}
\date{March 2012}
\maketitle

\begin{abstract}
This paper studies a simple dynamic linear panel regression model with
interactive fixed effects in which the variable of interest is measured with
error. To estimate the dynamic coefficient, we consider the least-squares
minimum distance (LS-MD) estimation method.


JEL\ Classification: C23, C26
\end{abstract}

\section{Introduction}

This paper studies a \textit{simple} dynamic linear panel regression model
with interactive fixed effects in which the variable of interest, say $%
Y_{it}^{\ast },$ contains measurement error: 
\begin{eqnarray}
Y_{it}^{\ast } &=&\alpha _{0}Y_{it-1}^{\ast }+\lambda
_{i}^{0}f_{t}^{0}+\epsilon _{it},\; \; \; \text{ }i=1,...,N,\; \; \text{ }%
t=1,...,T,  \label{model.unobserved} \\
Y_{it} &=&Y_{it}^{\ast }+\eta _{it}.  \label{model.classic.me}
\end{eqnarray}%
Here $Y_{it}$ is the observed variable and $\eta _{it}$ represents
measurement error. The term $\lambda _{i}^{0}f_{t}^{0}$ describes unobserved
interactive fixed effects.\footnote{%
In this paper, we consider a single factor, that is, the dimensions of $%
f_{t} $ and $\lambda _{i}$ are equal to one. The extension to the multiple
factor case is straightforward, but omitted due to space limitation.}$^{%
\text{,}}$\footnote{%
When interpreting $\lambda _{i}^{0}$ as individual specific fixed effects,
the term $f_{t}^{0}$ represents the (time-varying) linear projection
coefficient of $Y_{it}^{\ast }$ on $\lambda _{i}^{0}$ (holding $%
Y_{it-1}^{\ast }$ constant). This allows the effect of the unobserved
individual characteristic $\lambda _{i}^{0}$ on $Y_{it}^{\ast }$ to be
time-varying. Alternatively, one can interpret $f_{t}^{0}$ as a common time
specific shock (a common factor) and $\lambda _{i}^{0}$ then describes
reaction to the common shock (a factor loading).} The goal of the paper is
to estimate $\alpha _{0}$ when both the number of individuals $N$ and the
number of time periods $T$ are large.\footnote{%
We consider large $N,T$ approximations to characterize the bias due to the
incidental parameters $\lambda _{i}^{0}f_{t}^{0}$, see e.g. Bai (2009) and
Hahn and Kuersteiner (2004).}

The dynamics of the observed variable $Y_{it}$ can be written as 
\begin{equation}
Y_{it}=\alpha _{0}Y_{it-1}+\lambda _{i}^{0}f_{t}^{0}+U_{it},
\label{model.observed}
\end{equation}%
where $U_{it}=\epsilon _{it}+\eta _{it}-\alpha _{0}\eta _{it-1}.$ There are
two noticeable features in equations $\left( \ref{model.unobserved}\right) $
and $\left( \ref{model.observed}\right) $ compared to the widely studied
dynamic panel regression model. First, the individual effects take an
interactive form instead of the time invariant form. Secondly, the variable
of interest $Y_{it}^{\ast }$ is not observed but measured with error. To our
knowledge, combining these two features in dynamic linear panel regression
models has not been studied in the large $N,T$ panel literature.

We expect two hurdles in estimating $\alpha _{0}.$ One is the presence of
the interactive fixed effects $\lambda _{i}^{0}f_{t}^{0}$ which might cause
a so-called incidental parameter problem in both the cross section and the
time dimension. The second one is that the composite error $U_{it}$ in the
observed variable equation $\left( \ref{model.observed}\right) $ is
correlated with the lagged dependent variable $Y_{it-1}$ and we may
therefore need to use instrumental variables (IVs).

The main contribution of the paper is to find a valid estimation method that
overcomes these two problems. The proposed estimator is a nested two-step
estimator based on least squares minimization in the first step and distance
minimization for some of the first step parameter estimates in the second
step\footnote{%
An alternative approach would be to use the common correlated effect methods
suggested by Harding and Lamarche (2011). Both approaches have their own
merits and weaknesses. Comparing these different methods is not our interest
in this paper.}. Following Moon, Shum and Weidner (2012) (hereafter MSW), we
call this method the LS-MD estimation method. This approach was used in
estimating endogenous quantile regression models by Chernozhukov and Hansen
(2006, 2008) and in estimating the random coefficient logit demand model by
MSW.

\section{LS-MD Estimation}

The properties of the quasi-maximum likelihood estimator (QMLE), which
minimizes the sum of squared residuals, for large $N$, $T$ linear panel
regressions with interactive fixed effects were discussed in Bai (2009), and
Moon and Weidner (2010). However, this estimation method cannot be used to
estimate model $\left( \ref{model.observed}\right) $ since the regressor $%
Y_{it-1}$ is endogenous w.r.t. the error $U_{it}$ through the lagged
measurement error $\eta _{it-1}.$ In this case, we may use instrumental
variables. Since $U_{it}$ has an $MA(1)$ type serial dependence structure,
we have $E\left( U_{it}Y_{it-1-s}\right) =0$ for all $s\geq 1.$ This
suggests to choose $Z_{it}=\left( Z_{1,it},...,Z_{L,it}\right) ^{\prime
}=\left( Y_{it-2},...,Y_{it-1-L}\right) ^{\prime }$ for the IVs of the
endogenous regressor $Y_{it-1}$. The question, then, is how to use the
instrumental variables $Z_{it}$ to estimate $\alpha _{0}$ in the presences
of interactive fixed effects $\lambda _{i}^{0}f_{t}^{0}$ when both $N$ and $%
T $ are large.

The estimation method we consider in this paper is a two-step least-squares
minimum distance (LS-MD) estimation. This was recently proposed by MSW for
estimating the BLP demand model. A similar multi-step estimation idea was
also used in Chernozhukov and Hansen (2006, 2008) in estimating endogenous
quantile regressions with IVs.

The LS-MD estimation consists of the following two steps: Step 1: For given $%
\alpha ,$ we solve the least squares problem augmented by the instrumental
variables $Z_{it},$ that is, we run the OLS regression of $Y_{it}-\alpha
Y_{it-1}$ on $Z_{it}$ with interactive fixed effects $\lambda _{i}f_{t}$ and
solve%
\begin{equation*}
\left( \hat{\gamma}\left( \alpha \right) ,\text{ }\hat{\lambda}\left( \alpha
\right) ,\text{ }\hat{f}\left( \alpha \right) \right) =\arg \min_{\left(
\gamma ,\lambda ,f\right) }\sum_{i=1}^{N}\sum_{t=1}^{T}\left( Y_{it}-\alpha
Y_{it-1}-\gamma ^{\prime }Z_{it}-\lambda _{i}f_{t}\right) ^{2},
\end{equation*}%
where $\gamma =\left( \gamma _{1},...,\gamma _{L}\right) ^{\prime }$, $%
\lambda =\left( \lambda _{1},...,\lambda _{N}\right) ^{\prime }$ and $%
f=\left( f_{1},...,f_{T}\right) ^{\prime }.$ Step 2: For some positive
definite weight matrix $W_{NT}^{\gamma }$, we estimate $\alpha $ by
minimizing the length of $\hat{\gamma}\left( \alpha \right) $ as 
\begin{equation*}
\hat{\alpha}=\arg \min_{\alpha }\hat{\gamma}\left( \alpha \right) ^{\prime
}W_{NT}^{\gamma }\hat{\gamma}\left( \alpha \right) .
\end{equation*}

The idea of the LS-MD method is that since $Z_{it}$ is excluded in the
regression equation~$\left( \ref{model.observed}\right) $ the coefficient of 
$Z_{it}$ should be zero when $\alpha =\alpha _{0}$. When there is no
interactive fixed effect one can show that the LS-MD estimator is equivalent
to the conventional 2SLS estimator for an appropriate weight matrix $%
W_{NT}^{\gamma }$.

\section{Asymptotic Results}

\begin{assumption}
\label{a.consistency}(i) The unobserved error terms $\left \{ \epsilon
_{it}\right \} \sim iid\left( 0,\sigma _{\epsilon }^{2}\right) $ and $%
\left
\{ \eta _{it}\right \} \sim iid\left( 0,\sigma _{\eta }^{2}\right) $
across $i$ and over $t$ and $E\left \vert \epsilon _{it}\right \vert
^{\kappa },E\left \vert \eta _{it}\right \vert ^{\kappa }<\infty $ for some $%
\kappa >8.$ Also, $\left \{ \epsilon _{it}\right \} $ and $\left \{ \eta
_{it}\right \} $ are independent. (ii) Assume that $f_{t}^{0}$ are strictly
stationary and ergodic with $\sup_{t}\left \vert f_{t}\right \vert <\infty $
and $\frac{1}{T}\sum_{t=1}^{T}(f_{t}^{0})^{2}\rightarrow _{p}\Sigma _{f}>0,$
and $\lambda _{i}$ are iid with $\sup_{i}\left \vert \lambda
_{i}\right
\vert <\infty $ and $\frac{1}{N}\sum_{i=1}^{N}(\lambda
_{i}^{0})^{2}\rightarrow _{p}\Sigma _{\lambda }>0$. Also assume that $%
\left
\{ f_{t}^{0}\right \} ,\left \{ \lambda _{i}^{0}\right \} ,\left \{
\epsilon _{it}\right \} ,\left \{ \eta _{it}\right \} $ are independent.
(iii) $W_{NT}^{\gamma }\rightarrow _{p}W^{\gamma }>0.$ (vi) $\left \vert
\alpha _{0}\right \vert <1$ and $\alpha _{0}\neq 0.$
\end{assumption}

The iid assumptions of $\epsilon _{it}$ and $\eta _{it}$ are made for
simplicity of the analysis. Later, an extension to a non-iid case will be
discussed. Assumption \ref{a.consistency}(i) also assumes that the
measurement error $\eta _{it}$ is classical in the sense that $\eta _{it}$
has zero mean and is uncorrelated with $Y_{it}^{\ast }.$ Later we discuss
how to extend\ our method to some special cases of non-classical measurement
error. Assumption \ref{a.consistency}(ii) assumes that the factors are
strong, which is standard in the factor analysis literature. Assumption \ref%
{a.consistency}(vi) assumes that $\alpha _{0}\neq 0$, otherwise the IVs
become irrelevant.

Before we present the next assumption, we introduce some further notation.
We use $\left[ a_{it}\right] _{it}$ to denote an $N\times T$ matrix with
elements $a_{it}.$ For a full column rank matrix $A,$ let $\mathbb{P}%
_{A}=A\left( A^{\prime }A\right) ^{-1}A^{\prime }$ and $\mathbb{M}_{A}=I-%
\mathbb{P}_{A}.$ We use notation $Y=\left[ Y_{it}\right] _{it},$ $Y_{-k}=%
\left[ Y_{it-k}\right] _{it},$ $Z=\left[ Z_{it}\right] _{it},$ $U=\left[
U_{it}\right] _{it},$ $\epsilon =\left[ \epsilon _{it}\right] _{it},$ $\eta =%
\left[ \eta _{it}\right] _{it},$ and $\eta _{-1}=\left[ \eta _{it-1}\right]
_{it}.$ Define $\lambda ^{0}=\left( \lambda _{1}^{0},...,\lambda
_{N}^{0}\right) ^{\prime }$ and $f^{0}=\left( f_{1}^{0},...,f_{T}^{0}\right)
^{\prime }.$ We also define the $NT$-vectors $y_{-1}=vec\left( Y_{-1}\right) 
$ and $z=vec\left( Z\right) .$

\begin{assumption}
\label{a.relevance}Assume that there exists a positive constant $c>0$ such
that $\frac{1}{NT}y_{-1}^{\prime }\mathbb{P}_{z}y_{-1}- \max_{\lambda }\frac{%
1}{NT}y_{-1}^{\prime }\mathbb{P}_{I_{T}\otimes \tilde{\lambda}}y_{-1}>c$
with probability approaching one as $N,T \rightarrow \infty$, where $\tilde{%
\lambda}=\left( \lambda ^{0},\lambda \right) .$
\end{assumption}

Assumption \ref{a.relevance} is a relevance condition on the instruments. It
demands that the explanatory power of the instruments $Z_{it}$ for the
endogenous regressor $Y_{it-1}$, given by $\frac{1}{NT}y_{-1}^{\prime }%
\mathbb{P}_{z}y_{-1}$, is larger than the joint explanatory power for $%
Y_{it-1}$ of the true factor loading $\lambda ^{0}$ together with any other
factor loading $\lambda $, given by $\frac{1}{NT}y_{-1}^{\prime }\mathbb{P}%
_{I_{T}\otimes \tilde{\lambda}}y_{-1}$. If there are no interactive fixed
effects included in the model, then the assumption simplifies to the
standard relevance condition $\frac{1}{NT}y_{-1}^{\prime }\mathbb{P}%
_{z}y_{-1}>0$, which is satisfied for $\alpha ^{0}\neq 0$.

Suppose that Assumption~\ref{a.consistency} holds, and consider the special
case where $f_{t}^{0}$ has mean zero and is distributed independently over $%
t $. Then, Assumption \ref{a.relevance} is equivalent to\footnote{%
For the proof of this, we refer to the supplementary appendix which is
available at http://www.cemmap.ac.uk/publications.php.} 
\begin{equation}
\alpha _{0}^{2}>\frac{1+\frac{\sigma _{\epsilon }^{2}}{\Sigma _{\lambda
}\Sigma _{f}}+\frac{\sigma _{\eta }^{2}}{\Sigma _{\lambda }\Sigma _{f}}}{%
\left( 1+\frac{\sigma _{\epsilon }^{2}}{\Sigma _{\lambda }\Sigma _{f}}%
\right) ^{2}+\frac{\sigma _{\eta }^{2}}{\Sigma _{\lambda }\Sigma _{f}}}\;.
\label{suff_cond}
\end{equation}%
Thus, by imposing an appropriate lower bound on $|\alpha _{0}|,$ one can
guarantee that the lagged values of $Y_{it}$ are sufficiently relevant
instruments. The conclusion that an appropriate lower bound on $|\alpha
_{0}| $ is sufficient for the relevance assumption~Assumption \ref%
{a.relevance} can be extended to cases where $f_{t}^{0}$ is correlated
across $t$, but in general it is not possible to give such a convenient
analytic expression as in $\left( \ref{suff_cond}\right) $ for the lower
bound.\footnote{%
A non-zero mean of $f_{t}^{0}$ can result in situations where Assumption~\ref%
{a.relevance} is not satisfied for any value of $\alpha _{0}$. The
assumption that $f_{t}^{0}$ is mean zero would not be restrictive if we
would include a conventional individual specific fixed effect in the model,
in addition to the interactive fixed effect --- or equivalently (from an
asymptotic perspective), one can demean $Y_{it}$ separately for each $i$
before estimating the model with only interactive effects.} Note that the
lower bound in $\left( \ref{suff_cond}\right) $ goes to zero when $\Sigma
_{\lambda }\Sigma _{f}$ becomes small relative to $\sigma _{\epsilon }^{2}$,
i.e., the bound is not restrictive when the relative influence of the
factors on $Y_{it}$ is small.

\begin{theorem}
\label{t.consistency}Under Assumption \ref{a.consistency} and \ref%
{a.relevance} we have $\hat{\alpha}\rightarrow _{p}\alpha _{0}$ as $%
N,T\rightarrow \infty .\footnote{%
The proof is omitted due to space limitation. It is a special case of MSW
where their $\delta \left( \alpha \right) =Y-\alpha Y_{-1}$ and the
conditions in Assumptions \ref{a.consistency} and \ref{a.relevance} are
sufficient for the consistency conditions in MSW (see the supplementary
appendix available at http://www.cemmap.ac.uk/publications.php.)}^{\text{,}}%
\footnote{%
Note that Assumption \ref{a.relevance} is a sufficient condition for the
relevance of the instruments, but nothing is known about the necessity of
this assumption. The LS-MD estimator may also give consistent parameter
estimates in some situations where the assumption is violated.}$
\end{theorem}

To present the limiting distribution of $\hat{\alpha},$ we need to introduce
some further notation. Define the $NT$-vectors $y_{-1}^{\lambda f}$ and $%
z_{l}^{\lambda f}$ by $y_{-1}^{\lambda f}=vec\left( \mathbb{M}_{\lambda
^{0}}Y_{-1}\mathbb{M}_{f^{0}}\right)$, and $z_{l}^{\lambda f}=vec\left( 
\mathbb{M}_{\lambda ^{0}}Z_{l}\mathbb{M}_{f^{0}}\right)$, where $l=1,...,L.$
Let $u=vec\left( U\right) $ and $z^{\lambda f}=\left( z_{1}^{\lambda
f},...,z_{L}^{\lambda f}\right) .$

Define $G=\limfunc{plim}_{N,T \rightarrow \infty}\frac{1}{NT}y_{-1}^{\lambda
f\prime }z^{\lambda f}=\frac{\sigma _{\epsilon }^{2}}{1-\alpha _{0}^{2}}%
\left( \alpha _{0},\alpha_0^2,...,\alpha _{0}^{L}\right)^{\prime }$, and 
\begin{eqnarray*}
W &=&\limfunc{plim}_{N,T \rightarrow \infty}\left( \frac{1}{NT}z^{\lambda
f\prime }z^{\lambda f}\right) ^{-1}W_{NT}^{\gamma }\left( \frac{1}{NT}%
z^{\lambda f\prime }z^{\lambda f}\right) ^{-1} \\
&=&\left \{ \frac{\sigma _{\varepsilon }^{2}}{1-\alpha _{0}^{2}}\left[ 
\begin{array}{ccc}
1 & \cdots & \alpha _{0}^{L-1} \\ 
\vdots & \ddots & \vdots \\ 
\alpha _{0}^{L-1} & \cdots & 1%
\end{array}%
\right] +\sigma _{\eta }^{2}I_{L}\right \} W^{\gamma }\left \{ \frac{\sigma
_{\varepsilon }^{2}}{1-\alpha _{0}^{2}}\left[ 
\begin{array}{ccc}
1 & \cdots & \alpha _{0}^{L-1} \\ 
\vdots & \ddots & \vdots \\ 
\alpha _{0}^{L-1} & \cdots & 1%
\end{array}%
\right] +\sigma _{\eta }^{2}I_{L}\right \} .
\end{eqnarray*}%
Notice that under Assumption \ref{a.consistency}, the limits $G$ and $W$ are
well defined. Also, notice that under Assumption \ref{a.consistency}, we
have $GWG^{\prime }>0.$

Define 
\begin{equation*}
c^{\left( 2\right) }=\left[ C^{\left( 2\right) }\left( Z_{l},U\right) \right]
_{l=1,...,L},
\end{equation*}%
where 
\begin{align*}
& C^{\left( 2\right) }\left( Z_{l},U\right) \\
& =-\frac{1}{\sqrt{NT}}\left[ \limfunc{tr}\left( U\mathbb{M}%
_{f^{0}}U^{\prime }\mathbb{M}_{\lambda ^{0}}Z_{l}f^{0}\left( f^{0\prime
}f^{0}\right) ^{-1}\left( \lambda ^{0\prime }\lambda ^{0}\right)
^{-1}\lambda ^{0\prime }\right) +\limfunc{tr}\left( U^{\prime }\mathbb{M}%
_{\lambda ^{0}}U\mathbb{M}_{f^{0}}Z_{l}^{\prime }\lambda ^{0}\left( \lambda
^{0\prime }\lambda ^{0}\right) ^{-1}\left( f^{0\prime }f^{0}\right)
^{-1}f^{0\prime }\right) \right. \\
& \qquad \left. +\limfunc{tr}\left( U^{\prime }\mathbb{M}_{\lambda ^{0}}Z_{l}%
\mathbb{M}_{f^{0}}U^{\prime }\lambda ^{0}\left( \lambda ^{0\prime }\lambda
^{0}\right) ^{-1}\left( f^{0\prime }f^{0}\right) ^{-1}f^{0\prime }\right) %
\right] .
\end{align*}%
MSW showed that under Assumption \ref{a.consistency}, as $N,T\rightarrow
\infty $ with $\frac{N}{T}\rightarrow \kappa ^{2},$ where $0<\kappa <\infty
, $ we can approximate 
\begin{equation}
\sqrt{NT}\left( \hat{\alpha}-\alpha _{0}\right) =\left( GWG^{\prime }\right)
^{-1}GW\left[ \frac{1}{\sqrt{NT}}\left( z^{\lambda f}\right) ^{\prime
}u+c^{\left( 2\right) }\right] +o_{p}\left( 1\right) .  \label{appr.alphahat}
\end{equation}%
Notice that as $N,T\rightarrow \infty $ with $\frac{N}{T}\rightarrow \kappa
^{2},$ where $0<\kappa <\infty ,$ under Assumptions \ref{a.consistency} we
can show that%
\begin{equation}
\frac{1}{\sqrt{NT}}\left( z^{\lambda f}\right) ^{\prime }u+c^{\left(
2\right) }\Rightarrow N\left( -\kappa b,\Omega \right) ,
\label{limit.normality}
\end{equation}%
where $b=\left( b_{1},...,b_{L}\right) ^{\prime }$, and 
\begin{eqnarray*}
b_{l} &=&\limfunc{plim}_{N,T\rightarrow \infty }\frac{1}{N}\limfunc{tr}\left[
\mathbb{P}_{f^{0}}\left[ E\left( \epsilon ^{\prime }\tilde{\epsilon}%
_{-l-1}\right) +E\left( \left( \eta -\alpha _{0}\eta _{-1}\right) ^{\prime
}\eta _{-l-1}\right) \right] \right] \\
&&+\limfunc{plim}_{N,T\rightarrow \infty }\frac{1}{N}\limfunc{tr}\left[ 
\mathbb{E}(U^{\prime }U)\, \mathbb{M}_{f^{0}}\, \tilde{f}_{-l-1}^{0}\,(f^{0%
\prime }f^{0})^{-1}\,f^{0\prime }\right] , \\
\tilde{\epsilon}_{-l} &=&\left[ \tilde{\epsilon}_{it-l}\right] _{it},\quad 
\tilde{\epsilon}_{it-l}=\sum_{s=0}^{\infty }\alpha _{0}^{s}\epsilon
_{it-l-s},\quad \tilde{f}_{-l-1}^{0}=\left[ \sum_{s=0}^{\infty }\alpha
_{0}^{s}f_{t-1-l-s}\right] _{t}, \\
\Omega &=&\left( \sigma _{\epsilon }^{2}+\left( 1-\alpha _{0}\right)
^{2}\sigma _{\eta }^{2}\right) \left \{ \frac{\sigma _{\varepsilon }^{2}}{%
1-\alpha _{0}^{2}}\left[ 
\begin{array}{ccc}
1 & \cdots & \alpha _{0}^{L-1} \\ 
\vdots & \ddots & \vdots \\ 
\alpha _{0}^{L-1} & \cdots & 1%
\end{array}%
\right] +\sigma _{\eta }^{2}I_{L}\right \} .
\end{eqnarray*}%
\newline
Combining $\left( \ref{appr.alphahat}\right) $ and $\left( \ref%
{limit.normality}\right) ,$ we have the following theorem.

\begin{theorem}
\label{t.limit.normal}Suppose that Assumptions \ref{a.consistency} hold. As $%
N,T\rightarrow \infty $ with $\frac{N}{T}\rightarrow \kappa ^{2}$ and $%
0<\kappa <\infty ,$ we have 
\begin{equation*}
\sqrt{NT}\left( \hat{\alpha}-\alpha _{0}\right) \Rightarrow \mathcal{N}%
\left( -\kappa \left( GWG^{\prime }\right) ^{-1}GWb,\left( GWG^{\prime
}\right) ^{-1}GW\Omega WG^{\prime }\left( GWG^{\prime }\right) ^{-1}\right) .
\end{equation*}
\end{theorem}

Notice that the bias $b$ in the limit distribution is due to the incidental
parameters $\lambda _{i}^{0}f_{t}^{0}$ and the lagged dependent variables as
IVs, which is similar to the bias in Moon and Weidner (2010). This bias can
be consistently estimated and is correctable, for details we refer to Moon
and Weidner (2010) and Moon, Shum, and Weidner (2011).

\section{Monte Carlo Simulations}

In this section we investigate the finite sample properties of the LS-MD
estimator $\hat{\alpha}$ through small scale Monte Carlo simulations. The
data generating process is 
\begin{eqnarray*}
Y_{it}^{\ast } &=&\alpha _{0}Y_{it-1}^{\ast }+\lambda
_{i}^{0}f_{t}^{0}+\epsilon _{it}, \\
Y_{it} &=&Y_{it}^{\ast }+\eta _{it},
\end{eqnarray*}%
where $\alpha _{0}\in \left \{ 0.2,\text{ }0.5,\text{ }0.8\right \} ,$ $%
\left \{ \lambda _{i}\right \} ,\left \{ f_{t}\right \} ,\left \{ \eta
_{it}\right \} \sim iid$ $N\left( 0,0.4\right) $ and $\left \{ \epsilon
_{it}\right \} \sim iid$ $N\left( 0,1\right) .$ We consider various
combinations of $N\in \left \{ 20,50,100\right \} $ and $T\in \left \{
20,50,100\right \} .$ We use $Z_{it}=Y_{it-2}$ as an instrument. Notice that 
$\alpha _{0}=0.2$ violates the sufficient identification $\left( \ref%
{suff_cond}\right) $.%
\begin{equation*}
\text{Table 1. Monte Carlo Simulation Results}
\end{equation*}%
\medskip 
\begin{tabular}{c||ccc||ccc||ccc}
&  & ${\tiny \alpha }_{0}{\tiny =0.2}$ &  &  & ${\tiny \alpha }_{0}{\tiny %
=0.5}$ &  &  & ${\tiny \alpha }_{0}{\tiny =0.8}$ &  \\ \cline{2-10}
\textsc{N,T} & $\text{bias}$ & $\text{s.d.}$ & $\text{rmse}$ & $\text{bias}$
& $\text{s.d.}$ & $\text{rmse}$ & $\text{bias}$ & $\text{s.d.}$ & $\text{rmse%
}$ \\ \hline\hline
{\small 20,20} & {\small -0.173} & {\small 0.694} & {\small 0.715} & {\small %
-0.052} & {\small 0.171} & {\small 0.179} & {\small -0.030} & {\small 0.090}
& {\small 0.095} \\ 
{\small 20,50} & {\small -0.061} & {\small 0.292} & {\small 0.299} & {\small %
-0.004} & {\small 0.077} & {\small 0.077} & {\small -0.005} & {\small 0.031}
& {\small 0.032} \\ 
{\small 20,100} & {\small -0.005} & {\small 0.168} & {\small 0.168} & 
{\small -0.0004} & {\small 0.055} & {\small 0.055} & {\small -0.001} & 
{\small 0.021} & {\small 0.021} \\ 
{\small 50,20} & {\small -0.129} & {\small 0.440} & {\small 0.458} & {\small %
-0.022} & {\small 0.098} & {\small 0.100} & {\small -0.015} & {\small 0.061}
& {\small 0.063} \\ 
{\small 50,50} & {\small -0.012} & {\small 0.158} & {\small 0.158} & {\small %
-0.003} & {\small 0.048} & {\small 0.048} & {\small -0.001} & {\small 0.020}
& {\small 0.020} \\ 
{\small 50,100} & {\small -0.007} & {\small 0.102} & {\small 0.102} & 
{\small -0.001} & {\small 0.033} & {\small 0.033} & {\small -0.0005} & 
{\small 0.013} & {\small 0.013} \\ 
{\small 100,20} & {\small -0.092} & {\small 0.303} & {\small 0.316} & 
{\small -0.014} & {\small 0.068} & {\small 0.069} & {\small -0.014} & 
{\small 0.057} & {\small 0.059} \\ 
{\small 100,50} & {\small -0.008} & {\small 0.105} & {\small 0.105} & 
{\small -0.003} & {\small 0.034} & {\small 0.034} & {\small -0.001} & 
{\small 0.014} & {\small 0.014} \\ 
{\small 100,100} & {\small 0.001} & {\small 0.067} & {\small 0.067} & 
{\small -0.0001} & {\small 0.023} & {\small 0.023} & {\small -0.0003} & 
{\small 0.009} & {\small 0.009}%
\end{tabular}%
\newpage \noindent The finite sample properties of $\hat{\alpha}$, obtained
in simulations with 1000 repetitions, are reported in Table 1. Except for
the case of $\alpha _{0}=0.2$ with small samples, the LS-MD estimator $\hat{%
\alpha}$ performs well in finite samples.\footnote{%
We also investigated the finite sample properties of the bias corrected
estimator and found that analytical bias correction simultaneously reduces
the bias and the standard deviation of the estimator, except when both $%
\alpha ^{0}$ and $T$ are small. We omit the detailed results due to space
limitation, and since the biases in Table 1 without bias correction are
already quite small relative to the corresponding standard deviations.} When 
$\alpha _{0}=0.2,$ the finite sample properties improve as either $N$ and $T$
increases.

\section{Discussions}

\noindent \textbf{Choice of Instrumental Variables: }It is well known in the
GMM literature that the choice of moment conditions --- the choice of the
lag length $(L)$ in our setup --- is one of the important factors that
affect the finite sample properties of the GMM estimator. Various moment
condition selection procedures have been proposed in the literature. These
include, for example, the minimization of the (higher order) approximated
mean squared error (e.g., Donald and Newey (2001), Okui (2009), and
Kuersteiner (2010)) or of the asymptotic coverage error (e.g., Okui (2009)).
However, it is not straightforward to apply these procedures to the LS-MD
estimator. First, the LS-MD estimator has a bias even in the first order
approximation. Secondly, the key approximation techniques used in the
literature (e.g., Nagar's expansion and the Edgeworth expansion) are not
available in the exiting literature for the LS-MD estimator. Developing a
procedure for selection of $L$ is therefore beyond the scope of this paper.
\bigskip

\noindent \textbf{Extensions: }Our LS-MD estimation can be used for more
sophisticated cases. We briefly discuss how to extend our simple model.

\begin{enumerate}
\item \textbf{Inclusion of covariates}: The LS-MD estimation procedure can
be easily extended to include a model with other exogenous regressors, say $%
X_{it}.$ For example, in the first step one can regress $Y_{it}-\alpha
Y_{it-1}$ on $X_{it},$ $Z_{it}$ with interactive fixed effects $\lambda
_{i}f_{t}$ for fixed $\alpha .$ In the second step, minimize $\hat{\gamma}%
\left( \alpha \right) ^{\prime }W_{NT}^{\gamma }\hat{\gamma}\left( \alpha
\right) $ w.r.t. $\alpha .$

\item \textbf{Heteroskedastic error: }Until now, we assume that the errors $%
\epsilon _{it}$ and $\eta _{it}$ are homoskedastic for simplicity. If the
errors are heteroskedastic, then the term $c^{\left( 2\right) }$ contributes
additional bias terms to the limit distribution of $\hat{\alpha}$. These
biases are correctable (see e.g. Bai, 2009, and Moon and Weidner, 2010).%

\item \textbf{Non-classical measurement error}: Measurement error so far is
assumed to be classical. In many applications, however, measurement error
can be correlated with the unobserved latent variable and the covariates.
Our estimation method is still valid under more general measurement error
models. For example, suppose that people tend to report income $Y_{it}$
proportionally to $Y_{it}^{\ast }$ as 
\begin{equation}
Y_{it}=\gamma _{0i}+\gamma _{1it}Y_{it}^{\ast }+v_{it},\text{ }
\label{measurement.error}
\end{equation}%
where $v_{it}$ is an unobserved error. Note that the measurement error in
model~$\left( \ref{measurement.error}\right) $ is non-classical since the
measurement error, $\eta _{it}=Y_{it}-Y_{it}^{\ast },$ could be correlated
with $Y_{it}^{\ast }$ and the mean of the measurement error is not
necessarily zero.\footnote{%
A special case of model $\left( \ref{measurement.error}\right) $ is $\gamma
_{0it}=0$ and $\gamma _{1it}=1,$ in which case $v_{it}$ is classical.} Model 
$\left( \ref{measurement.error}\right) $ is a modified version of a linear
measurement error model that allows for a heterogeneous relationship between 
$Y_{it}$ and $Y_{it}^{\ast }$ across cross-section and over time.\footnote{%
Bollinger and Chandra (2005) and Kim and Solon (2005) developed a model
allowing for a constant linear relationship between $Y_{it}\ $and $%
Y_{it}^{\ast }$, based on the evidence in surveyed income; i.e., those who
earn higher than average tend to report their earning less, while those who
earn lower than average tend to report higher. See also Bound, Brown and
Mathiowetz (2001).} When the coefficient $\gamma _{1it}$ is random
satisfying $\gamma _{1it}=\gamma _{1}+w_{it},$ where $\left \{ w_{it}\right
\} $ and $\left \{ v_{it}\right \} $ are $iid$ across $i$ and over $t$ with
zero mean, and $\left \{ w_{it}\right \} ,\left \{ v_{it}\right \} ,\left \{
\epsilon _{it}\right \} $ are independent of each other, then we have the
following dynamic equation with two factors (or one factor and a time
invariant fixed effect) as 
\begin{subequations}
\begin{equation*}
Y_{it}=\alpha Y_{it-1}+\delta _{i}^{\prime }h_{t}+U_{it},\text{ }
\end{equation*}%
where $\delta _{i}=\left( \gamma _{1}\lambda _{i},\left( 1-\alpha \right)
\gamma _{0i},\right) ^{\prime },$ $h_{t}=\left( f_{t},1\right) ^{\prime }$
and 
\end{subequations}
\begin{equation}
U_{it}=\gamma _{1}\epsilon _{it}+v_{it}-\alpha v_{it-1}+Y_{it}^{\ast
}w_{it}-\alpha Y_{it-1}^{\ast }w_{it-1}.  \label{composite.error.2}
\end{equation}%
Note that the composite error $U_{it}$ in $\left( \ref{composite.error.2}%
\right) $ has serial dependence structure similar to an $MA\left( 1\right) $
process, and in this case $Z_{it}=\left( Z_{1,it},...,Z_{L,it}\right)
^{\prime }=\left( Y_{it-2},...,Y_{it-1-L}\right) ^{\prime }$ still remains
uncorrelated with $U_{it}.$ \pagebreak
\end{enumerate}

\pagebreak

\section{Supplementary Appendix (Not for Publication)}

\subsection{Proof of Consistency}

We show that Assumptions 3.1 and 3.2 in the current model are sufficient for
Assumption~1 of Moon, Shum, and Weidner (2012) (MSW hereafter) with $\delta
\left( \alpha \right) $ in MSW replaced by $Y-\alpha Y_{-1}$, and $X_{k}$ in
MSW replaced by $0$.

Notation: When $A$ is a matrix, $\left \Vert A\right \Vert ^{2}$ denotes the
largest eigenvalue of $A^{\prime }A$ and $\left \Vert A\right \Vert _{F}^{2}$
denotes the trace of $A^{\prime }A.$

\begin{itemize}
\item Assumption 1(i) holds since uniformly in $\alpha $ outside of any
neighborhood of $\alpha _{0}$ we have%
\begin{equation*}
\frac{\left \Vert \delta \left( \alpha \right) -\delta \left( \alpha
_{0}\right) \right \Vert _{F}}{\left \Vert \alpha -\alpha _{0}\right \Vert }%
=\left \Vert Y_{-1}\right \Vert _{F}=\sqrt{\sum_{i=1}^{N}%
\sum_{t=1}^{T}Y_{it-1}^{2}}=O_{p}\left( \sqrt{NT}\right) .
\end{equation*}%
Also, it follows that 
\begin{equation*}
\left \Vert Z_{l}\right \Vert _{F}=\sqrt{\sum_{i=1}^{N}%
\sum_{t=1}^{T}Y_{it-1-l}^{2}}=O_{p}\left( \sqrt{NT}\right) .
\end{equation*}

\item Assumption 1(ii) is satisfied because $\left \Vert U\right \Vert
=\left \Vert \epsilon +\eta -\alpha _{0}\eta _{-1}\right \Vert \leq
\left
\Vert \epsilon \right \Vert +\left \Vert \eta \right \Vert
+\left
\vert \alpha _{0}\right \vert \left \Vert \eta _{-1}\right \Vert
=O_{p}\left( \sqrt{\max \left( J,T\right) }\right) $ because $\left \{
\epsilon _{it}\right \} ,\left \{ \eta _{it}\right \} \sim iid$ with mean
zero and finite moments higher than 4 (See Moon and Weidner (2010)).

\item Assumption 1(iii)%
\begin{equation*}
\frac{1}{NT}\sum_{i=1}^{N}\sum_{t=1}^{T}Y_{it-1-l}U_{it}=o_{p}\left( 1\right)
\end{equation*}%
follows for $l\geq 1$ since $E\left( Y_{it-1-l}U_{it}\right) =0$ if $l\geq
1. $

\item Assumption 1(iv) follows since any (nontrivial) linear combinations of 
$Z_{l}^{\prime }s$ have rank higher than two under Assumption 3.1.

\item Assumption 1(v) holds by Assumption 3.2 with $\Delta \xi _{\alpha
,\beta }=-\left( \alpha -\alpha _{0}\right) y_{-1}.$

\item Assumption 1(vi) holds by Assumption 3.1 (iii).
\end{itemize}

\subsection{Asymptotic Normality}

\begin{itemize}
\item Assumptions 2 and 3 in MSW follow immediately under Assumption 3.1.

\item Assumptions 4(i),(ii) and 5 in MSW follow since in this paper $\delta
\left( \alpha \right) =Y-\alpha Y_{-1}$ is linear in $\alpha $ and by the
conditions in Assumption 3.1.

\item Assumption 4(iv) of MSW is satisfied with 
\begin{eqnarray*}
Z_{l,it}^{\text{str}} &=&\lambda _{i}\sum_{s=0}^{\infty }\alpha
_{0}^{s}f_{t-1-l-s} \\
Z_{l,it}^{\text{weak}} &=&\sum_{s=0}^{\infty }\alpha _{0}^{s}\epsilon
_{it-1-l-s}+\eta _{it-1-l}.
\end{eqnarray*}

\item Notice that the conditions in Assumption 4(iii) of MSW are satisfied
except for that $U_{it}$ is an MA(1) type error over time, that is, $U_{it}$
and $U_{it-1}$ are dependent, while $U_{it}$ and $U_{it-s}$ are independent
for $s\geq 2.$ Because of this, we need to modify the proof of Theorem 5.2
of MSW and in what follows we give a sketch.

\item Step 1: First we show that 
\begin{equation*}
\sqrt{N}\left( \hat{\alpha}-\alpha _{0}\right) =O_{p}\left( 1\right) .
\end{equation*}

\item Step 2: Using the asymptotic likelihood exansion derived in Moon and
Weidner (2010), we can approximate $\sqrt{NT}\hat{\gamma}\left( \alpha
\right) $ as a linear function of $\sqrt{NT}\left( \alpha -\alpha
_{0}\right) ;$ 
\begin{equation*}
\sqrt{NT}\hat{\gamma}\left( \alpha \right) =\left( \frac{1}{NT}z^{\lambda
f\prime }z^{\lambda f}\right) ^{-1}\left[ \frac{1}{\sqrt{NT}}z^{\lambda
f\prime }u+c^{\left( 2\right) }-z^{\lambda f\prime }y_{-1}^{\lambda f}\sqrt{%
NT}\left( \alpha -\alpha _{0}\right) \right] +o_{p}\left( 1\right)
\end{equation*}%
where $o_{p}\left( 1\right) $ holds uniformly in $\alpha $ with $\sqrt{N}%
\left \vert \alpha -\alpha _{0}\right \vert <c$ for all $c.$

\item Step 3: We then approximate the second step objective function as a
quadratic function of $\sqrt{NT}\left( \alpha -\alpha _{0}\right) $ by
plugging the linear approximation of $\sqrt{NT}\hat{\gamma}\left( \alpha
\right) .$ Then, we deduce that 
\begin{eqnarray*}
&&\sqrt{NT}\left( \hat{\alpha}-\alpha _{0}\right) \\
&=&\left[ \left( \frac{1}{NT}y_{-1}^{\lambda f\prime }z^{\lambda f}\right)
\left( \frac{1}{NT}z^{\lambda f\prime }z^{\lambda f}\right)
^{-1}W_{NT}^{\gamma }\left( \frac{1}{NT}z^{\lambda f\prime }z^{\lambda
f}\right) ^{-1}\left( \frac{1}{NT}z^{\lambda f\prime }y_{-1}^{\lambda
f}\right) \right] ^{-1} \\
&&\times \left( \frac{1}{NT}y_{-1}^{\lambda f\prime }z^{\lambda f}\right)
\left( \frac{1}{NT}z^{\lambda f\prime }z^{\lambda f}\right)
^{-1}W_{NT}^{\gamma }\left( \frac{1}{NT}z^{\lambda f\prime }z^{\lambda
f}\right) ^{-1}\left[ \frac{1}{\sqrt{NT}}z^{\lambda f\prime }u+c^{\left(
2\right) }\right] +o_{p}\left( 1\right) \\
&=&\left( GWG^{\prime }\right) ^{-1}GW\left( \frac{1}{\sqrt{NT}}z^{\lambda
f\prime }u+c^{\left( 2\right) }\right) +o_{p}\left( 1\right) ,
\end{eqnarray*}%
as required for $\left( \ref{appr.alphahat}\right) .$ (Notice that Steps
1,2,3 are not affected by the MA(1) type dependence of $U_{it}.)$

\item Step 4: By definition%
\begin{eqnarray*}
\frac{1}{\sqrt{NT}}z^{\lambda f\prime }u &=&\left[ 
\begin{array}{c}
\frac{1}{\sqrt{NT}}\mathrm{tr}\left( \mathbb{M}_{f^{0}}U^{\prime }\mathbb{M}%
_{\lambda ^{0}}Z_{1}\right) \\ 
\vdots \\ 
\frac{1}{\sqrt{NT}}\mathrm{tr}\left( \mathbb{M}_{f^{0}}U^{\prime }\mathbb{M}%
_{\lambda ^{0}}Z_{L}\right)%
\end{array}%
\right] =\left[ 
\begin{array}{c}
\frac{1}{\sqrt{NT}}\mathrm{tr}\left( \mathbb{M}_{f^{0}}U^{\prime }\mathbb{M}%
_{\lambda ^{0}}Z_{1}^{\text{weak}}\right) \\ 
\vdots \\ 
\frac{1}{\sqrt{NT}}\mathrm{tr}\left( \mathbb{M}_{f^{0}}U^{\prime }\mathbb{M}%
_{\lambda ^{0}}Z_{L}^{\text{weak}}\right)%
\end{array}%
\right] \\
&=&\frac{1}{\sqrt{NT}}\sum_{i=1}^{T}\sum_{t=1}^{T}U_{it}Z_{it}^{\text{weak}}-%
\left[ \frac{1}{\sqrt{NT}}\mathrm{tr}\left( \mathbb{P}_{f^{0}}E\left(
U^{\prime }Z_{l}^{\text{weak}}\right) \right) \right] _{l=1,...,L} \\
&&-\left[ \frac{1}{\sqrt{NT}}\mathrm{tr}\left( \mathbb{P}_{f^{0}}\left(
U^{\prime }Z_{l}^{\text{weak}}-E\left( U^{\prime }Z_{l}^{\text{weak}}\right)
\right) \right) \right] _{l=1,...L} \\
&&+\left[ \frac{1}{\sqrt{NT}}\mathrm{tr}\left( U^{\prime }\mathbb{P}%
_{\lambda ^{0}}Z_{l}^{\text{weak}}\right) \right] _{l=1,...,L}+\left[ \frac{1%
}{\sqrt{NT}}\mathrm{tr}\left( \mathbb{P}_{f^{0}}U^{\prime }\mathbb{P}%
_{\lambda ^{0}}Z_{l}^{\text{weak}}\right) \right] _{l=1,...,L} \\
&=&I+II+III+IV+V,\text{ say.}
\end{eqnarray*}%
Then, by the CLT (e.g., Moon and Phillips (1999)) we have 
\begin{equation*}
\frac{1}{\sqrt{NT}}\sum_{i=1}^{T}\sum_{t=1}^{T}U_{it}Z_{it}^{\text{weak}%
}\Rightarrow N\left( 0,\Omega \right) ,
\end{equation*}%
where%
\begin{eqnarray*}
\Omega &=&E\left( U_{it}^{2}\left( Z_{it}^{\text{weak}}\right) \left(
Z_{it}^{\text{weak}}\right) ^{\prime }\right) +E\left( U_{it}U_{it-1}\left(
Z_{it}^{\text{weak}}\right) \left( Z_{it-1}^{\text{weak}}\right) ^{\prime
}\right) \\
&&+E\left( U_{it}U_{it+1}\left( Z_{it}^{\text{weak}}\right) \left( Z_{it+1}^{%
\text{weak}}\right) ^{\prime }\right) .
\end{eqnarray*}%
A direct calculation shows that 
\begin{eqnarray*}
&&E\left( U_{it}^{2}\left( Z_{it}^{\text{weak}}\right) \left( Z_{it}^{\text{%
weak}}\right) ^{\prime }\right) \\
&=&E\left( U_{it}^{2}\right) E\left[ \left( Z_{it}^{\text{weak}}\right)
\left( Z_{it}^{\text{weak}}\right) ^{\prime }\right] \\
&=&\left( \sigma _{\epsilon }^{2}+\left( 1+\alpha _{0}^{2}\right) \sigma
_{\eta }^{2}\right) \left \{ \frac{\sigma _{\varepsilon }^{2}}{1-\alpha
_{0}^{2}}\left[ 
\begin{array}{ccc}
1 & \cdots & \alpha _{0}^{L-1} \\ 
\vdots & \ddots & \vdots \\ 
\alpha _{0}^{L-1} & \cdots & 1%
\end{array}%
\right] +\sigma _{\eta }^{2}I_{L}\right \}
\end{eqnarray*}%
and%
\begin{eqnarray*}
&&E\left( U_{it}U_{it-1}\left( Z_{it}^{\text{weak}}\right) \left( Z_{it-1}^{%
\text{weak}}\right) ^{\prime }\right) +E\left( U_{it}U_{it+1}\left( Z_{it}^{%
\text{weak}}\right) \left( Z_{it+1}^{\text{weak}}\right) ^{\prime }\right) \\
&=&-2\alpha _{0}\sigma _{\eta }^{2}\left \{ \frac{\sigma _{\varepsilon }^{2}%
}{1-\alpha _{0}^{2}}\left[ 
\begin{array}{ccc}
1 & \cdots & \alpha _{0}^{L-1} \\ 
\vdots & \ddots & \vdots \\ 
\alpha _{0}^{L-1} & \cdots & 1%
\end{array}%
\right] +\sigma _{\eta }^{2}I_{L}\right \} ,
\end{eqnarray*}%
which leads 
\begin{equation*}
\Omega =\left( \sigma _{\epsilon }^{2}+\left( 1-\alpha _{0}\right)
^{2}\sigma _{\eta }^{2}\right) \left \{ \frac{\sigma _{\varepsilon }^{2}}{%
1-\alpha _{0}^{2}}\left[ 
\begin{array}{ccc}
1 & \cdots & \alpha _{0}^{L-1} \\ 
\vdots & \ddots & \vdots \\ 
\alpha _{0}^{L-1} & \cdots & 1%
\end{array}%
\right] +\sigma _{\eta }^{2}I_{L}\right \}
\end{equation*}

Also, a direct calculation shows that 
\begin{eqnarray*}
&&-\limfunc{plim}_{N,T\rightarrow \infty }\frac{1}{\sqrt{NT}}\mathrm{tr}%
\left( \mathbb{P}_{f^{0}}E\left( U^{\prime }Z_{l}^{\text{weak}}\right)
\right) \\
&=&-\kappa \limfunc{plim}_{N,T\rightarrow \infty }\frac{1}{N}\limfunc{tr}%
\left[ \mathbb{P}_{f^{0}}\left[ E\left( \epsilon ^{\prime }\tilde{\epsilon}%
_{-l-1}\right) +E\left( \left( \eta -\alpha _{0}\eta _{-1}\right) ^{\prime
}\eta _{-l-1}\right) \right] \right] .
\end{eqnarray*}

By modifying Lemma C.2 (f), (j), and (m) in Moon and Weidner (2010), we can
show that 
\begin{equation*}
III,IV,V=o_{p}\left( 1\right) .
\end{equation*}

\item Step 5: By modifying Lemma C.2 (c), (d), (e), (g), (h), (i), (k), and
(l) in Moon and Weidner (2010), we can show that%
\begin{equation*}
\limfunc{plim}_{N,T\rightarrow \infty }c^{\left( 2\right) }=-\kappa \left[ 
\limfunc{plim}_{N,T\rightarrow \infty }\frac{1}{N}\limfunc{tr}\left[ \mathbb{%
E}(U^{\prime }U)\, \mathbb{M}_{f^{0}}\, \tilde{f}_{-l-1}^{0}\,(f^{0\prime
}f^{0})^{-1}\,f^{0\prime }\right] \right] _{l=1,...,L}.
\end{equation*}

\item Step 6: Combining the limits in Steps 4 and 5 yields the desired
result in $\left( \ref{limit.normality}\right) .$
\end{itemize}

\subsection{Sufficient Conditions for Assumption~\protect\ref{a.relevance}}

In matrix notation we can write $\left( \ref{model.observed}\right) $ as 
\begin{equation*}
Y=\alpha ^{0}Y_{-1}+\lambda ^{0}f^{0\prime }+U.
\end{equation*}%
By recursively applying the model we find 
\begin{equation*}
Y=\lambda ^{0}F^{0\prime }+E+Y^{\mathrm{init}},
\end{equation*}%
where $F$ is the $T\times 1$ vector with entries $F_{t}=\sum_{\tau
=0}^{t-1}\alpha _{0}^{\tau }f_{t-\tau }^{0}$, and $E$ and $Y^{\mathrm{init}}$
are the $T\times N$ matrices with entries $E_{it}=\eta _{it}+\sum_{\tau
=0}^{t-1}\alpha _{0}^{\tau }\epsilon _{t-\tau }$, and $Y_{it}^{\mathrm{init}%
}=\alpha _{0}^{t}Y_{i0}$. We denote lagged versions of $F^{0}$ and $E$ by $%
F_{-1}^{0}$ and $E_{-1}$, etc.

In the following we assume $L=1$. In that case Assumption~\ref{a.relevance}
is satisfied if 
\begin{equation}
\frac{\left( \limfunc{plim}_{N,T\rightarrow \infty }\frac{1}{NT}%
y_{-1}^{\prime }z\right) ^{2}}{\limfunc{plim}_{N,T\rightarrow \infty }\frac{1%
}{NT}z^{\prime }z}-\limfunc{plim}_{N,T\rightarrow \infty }\left(
\max_{\lambda }\frac{1}{NT}y_{-1}^{\prime }\mathbb{P}_{I_{T}\otimes \tilde{%
\lambda}}y_{-1}\right) >0,  \label{relevance2}
\end{equation}%
where $\tilde{\lambda}=\left( \lambda ^{0},\lambda \right) $. If $f_{t}$ is
mean zero and independent across $t$ we have 
\begin{align*}
\frac{1}{NT}y_{-1}^{\prime }z& =\frac{1}{NT}\mathrm{tr}(Y_{-1}Y_{-2}) \\
& =\frac{1}{NT}\mathrm{tr}(E_{-1}E_{-2})+\frac{1}{NT}\Vert \lambda \Vert
^{2}(F_{-1}^{\prime }F_{-2})+o_{p}(1) \\
& =\frac{\alpha _{0}}{1-\alpha _{0}^{2}}\left( \sigma _{\epsilon
}^{2}+\Sigma _{\lambda }\Sigma _{f}\right) +o_{p}(1), \\
\frac{1}{NT}z^{\prime }z& =\frac{1}{NT}\mathrm{tr}(Y_{-1}Y_{-1}) \\
& =\frac{1}{1-\alpha _{0}^{2}}\left( \sigma _{\epsilon }^{2}+\Sigma
_{\lambda }\Sigma _{f}\right) +\sigma _{\eta }^{2}+o_{p}(1),
\end{align*}%
and 
\begin{align*}
\max_{\lambda }\frac{1}{NT}y_{-1}^{\prime }\mathbb{P}_{I_{T}\otimes \tilde{%
\lambda}}y_{-1}& =\max_{\lambda }\frac{1}{NT}\mathrm{tr}(Y_{-1}^{\prime }%
\mathbb{P}_{\tilde{\lambda}}Y_{-1}) \\
& =\max_{\lambda }\frac{1}{NT}\mathrm{tr}(F_{-1}\lambda ^{\prime }\mathbb{P}%
_{\tilde{\lambda}}\lambda F_{-1}^{\prime })+o_{p}(1) \\
& =\frac{1}{NT}\Vert \lambda \Vert ^{2}\Vert F_{-1}\Vert ^{2}+o_{p}(1) \\
& =\frac{1}{1-\alpha _{0}^{2}}\Sigma _{\lambda }\Sigma _{f}+o_{p}(1).
\end{align*}%
Plugging these results into condition \eqref{relevance2} yields condition %
\eqref{suff_cond}.

\end{document}